\documentclass[aps,prd,twocolumn,superscriptaddress,showpacs,amsmath,amssymb,
preprintnumbers,nofootinbib, 10pt]{revtex4-2}
\usepackage{graphicx} 
\usepackage{epstopdf} 
\usepackage{booktabs}
\usepackage{dcolumn} 
\usepackage{xcolor} 
\usepackage{amsmath,amssymb} 
\usepackage{hyperref} 
\usepackage[utf8]{inputenc} 
\usepackage[english]{babel} 
\usepackage[mathlines]{lineno} 
\usepackage{blindtext} 
\usepackage{ifthen} 
\usepackage{etoolbox}
\usepackage{orcidlink} 
\usepackage{slashed}
\graphicspath{{figures/}} 

\newboolean{articletitles}
\setboolean{articletitles}{true} 

\RequirePackage[normalem]{ulem} 
\RequirePackage[normalem]{ulem} 
\RequirePackage{color}\definecolor{RED}{rgb}{1,0,0}\definecolor{BLUE}{rgb}{0,0,1} 
\providecommand{\DIFadd}[1]{{\protect\color{blue}\uwave{#1}}} 
\providecommand{\DIFdel}[1]{{\protect\color{red}\sout{#1}}}                      

\input{belle2-symbols}

\newcommand{\ifb}{\ensuremath{{\rm fb}^{-1}}\xspace}
\newcommand{\lumi}{\ensuremath{365~\ifb}\xspace}
\newcommand{\NBB}{\ensuremath{N_{\Upsilon(4S)}=(387 \pm 6)\times 10^6}\xspace}
\newcommand{\NBBval}{\ensuremath{(387 \pm 6)\times 10^6}\xspace}
\newcommand{\resVcbCLN}{\ensuremath{(38.5\pm 1.3)\times 10^{-3}}\xspace}
\newcommand{\resVcbBCL}{\ensuremath{(39.2\pm 0.4\,(\mathrm{stat.}) \pm 0.6\,(\mathrm{sys.}) \pm 0.5\,(\mathrm{th.}) )\times 10^{-3}}\xspace}
\newcommand{\resBrBz}{\ensuremath{(2.06 \pm 0.05\,(\mathrm{stat.}) \pm 0.10\,(\mathrm{sys.}))\%}\xspace}
\newcommand{\resBrB}{\ensuremath{(2.31 \pm 0.04\,(\mathrm{stat.}) \pm 0.09\,(\mathrm{sys.}))\%}\xspace}
\newcommand{\resRemu}{\ensuremath{1.020 \pm 0.020\,(\mathrm{stat.})\pm 0.022\,(\mathrm{sys.})
}\xspace}

\newcommand{\dGidw}{\ensuremath{\Delta \Gamma_i / \Delta w}}

\begin{document}
\begin{titlepage}
\includegraphics[width = 3cm]{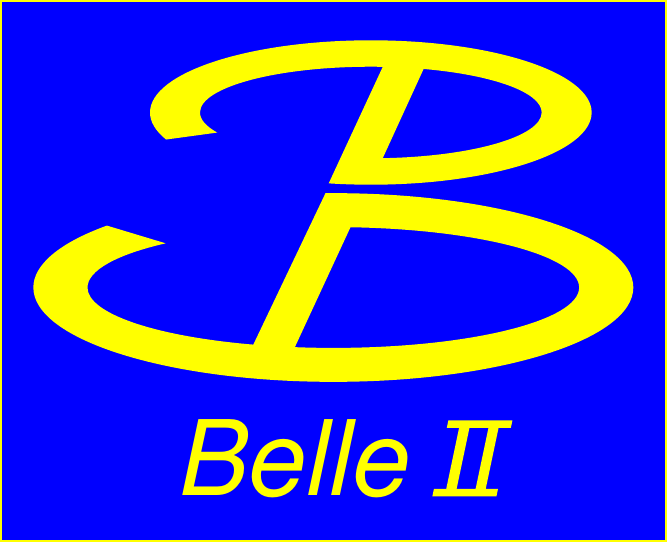}\vspace*{-3cm}

\begin{flushright}
\vspace*{5ex}
Belle II Preprint 2025-004\\
KEK Preprint 2025-1\\
~
\end{flushright}
\vspace*{36pt}

\title{Determination of $|V_{cb}|$ using $B\to D\ell\nu_\ell$ Decays at Belle II
}
  \author{I.~Adachi\,\orcidlink{0000-0003-2287-0173}} 
  \author{K.~Adamczyk\,\orcidlink{0000-0001-6208-0876}} 
  \author{L.~Aggarwal\,\orcidlink{0000-0002-0909-7537}} 
  \author{H.~Ahmed\,\orcidlink{0000-0003-3976-7498}} 
  \author{Y.~Ahn\,\orcidlink{0000-0001-6820-0576}} 
  \author{H.~Aihara\,\orcidlink{0000-0002-1907-5964}} 
  \author{N.~Akopov\,\orcidlink{0000-0002-4425-2096}} 
  \author{S.~Alghamdi\,\orcidlink{0000-0001-7609-112X}} 
  \author{M.~Alhakami\,\orcidlink{0000-0002-2234-8628}} 
  \author{A.~Aloisio\,\orcidlink{0000-0002-3883-6693}} 
% ...
\collaboration{The Belle II Collaboration}

\begin{abstract}
We present a determination of the Cabibbo-Kobayashi-Maskawa matrix element $|V_{cb}|$ from the decay $B\to D\ell\nu_\ell$ using a \lumi $e^+e^-\to\Upsilon(4S)\to B\bar B$ data sample recorded by the Belle~II experiment at the SuperKEKB collider. The semileptonic decay of one $B$~meson is reconstructed in the modes $B^0\to D^-(\to K^+\pi^-\pi^-)\ell^+\nu_\ell$ and $B^+\to \bar D^0(\to K^+\pi^-)\ell^+\nu_\ell$, where $\ell$ denotes either an electron or a muon. Charge conjugation is implied. The second $B$~meson in the $\Upsilon(4S)$~event is not reconstructed explicitly. Using an inclusive reconstruction of the unobserved neutrino momentum, we determine the recoil variable $w=v_B\cdot v_D$, where $v_B$ and $v_D$ are the 4-velocities of the $B$ and $D$~mesons. We measure the total decay branching fractions to be $\mathcal{B}(B^0\to D^-\ell^+\nu_\ell)=\resBrBz$ and $\mathcal{B}(B^+\to\bar D^0\ell^+\nu_\ell)=\resBrB$. We probe lepton flavor universality by measuring $\mathcal{B}(B\to De\nu_e)/\mathcal{B}(B\to D\mu\nu_\mu)=\resRemu$. Fitting the partial decay branching fraction as a function of $w$ and using the average of lattice QCD calculations of the $B\to D$~form factor, we obtain $ |V_{cb}|=\resVcbBCL$.
\end{abstract}

\maketitle
\end{titlepage}

\section{Introduction}

In the Standard Model (SM) of particle physics, the Cabibbo-Kobayashi-Maskawa (CKM) matrix describes the relationship between the weak interaction and the mass eigenstates of quarks~\cite{Cabibbo:1963yz, Kobayashi:1973fv}. The squared magnitude of the matrix element $V_{cb}$ determines the transition rate of $b$ into $c$ quarks and must be determined experimentally. The precise knowledge of this fundamental parameter of the SM is important for the ongoing precision flavor physics program at the Belle II experiment and elsewhere, and probes the mechanism of quark flavor mixing in the SM. The CKM magnitude $|V_{cb}|$ can be measured from inclusive semileptonic decays $B\to X_c\ell\nu_\ell$, where $X_c$ is a hadronic system with a charm quark and $\ell$ is an electron or muon, or determinations can be based on a single, exclusive mode such as $B\to D^*\ell\nu_\ell$ or $B\to D\ell\nu_\ell$. These two approaches currently differ by about three standard deviations~\cite{HeavyFlavorAveragingGroupHFLAV:2024ctg} motivating further research.

For the measurement of $|V_{cb}|$, the decay $B\to D\ell\nu_\ell$ has received less attention than $B\to D^*\ell\nu_\ell$,
with few recent measurements available~\cite{BaBar:2009zxk,Belle:2015pkj,babarcollaboration2023modelindependentextractionformfactors}. This is largely due to its smaller branching fraction and significant experimental backgrounds, particularly from $B\to D^*\ell\nu_\ell$ feed-down. These disadvantages can be mitigated with the large data samples
available at the $B$ factory experiments, in which case the advantages of $B\to D\ell\nu_\ell$ become clear: the $B\to D$~transition can be described by the single vector form factor $f_+(q^2)$ and thus has smaller theoretical uncertainties than $B\to D^* \ell\nu_\ell$. The reconstruction of $B\to D\ell\nu_\ell$ does not require the measurement of the slow pion from the $D^*$~decay, which is the leading experimental uncertainty in recent measurements of $B\to D^*\ell\nu_\ell$~\cite{Belle-II:2023okj}. In addition, in $B\to D\ell\nu_\ell$ both isospin states are accessible with similar experimental precision, allowing us to probe isospin violating contributions resulting from long-distance radiative corrections to $B^0\to D^-\ell^+\nu_\ell$ decays. In contrast to previous comparable measurements by the BaBar and Belle collaborations~\cite{BaBar:2009zxk,babarcollaboration2023modelindependentextractionformfactors,Belle:2015pkj}, which fully reconstructed the second $B$ meson in the event hadronically, this analysis reconstructs the tag side inclusively from the remaining particles in the event. This approach is similar to the inclusive tagging approach employed when searching for $B\to K\nu\bar\nu$ at Belle II~\cite{knunu}, and allows for higher signal yields.


This paper is organized as follows: After reviewing the theory of the decay $B\to D\ell\nu_\ell$ in Sec.~\ref{sec:theo}, we describe the experimental procedure in Sec.~\ref{sec:exp}. Our results are presented in Sec.~\ref{sec:res} together with a detailed discussion of experimental systematic uncertainties. Finally, Sec.~\ref{sec:vcb} converts our measurement of the decay $B\to D\ell\nu_\ell$ into a determination of the CKM magnitude $|V_{cb}|$. Throughout this paper charge conjugation is implied. We refer to $B^0\to D^-\ell^+\nu_\ell$ as the \textit{neutral mode}, and to $B^+\to\bar D^0\ell^+\nu_\ell$ as the \textit{charged mode} in reference to the $B$ meson charge.

\section{Theoretical framework} \label{sec:theo}

The differential decay width of $B\to D\ell\nu_\ell$ can be expressed as a function of $w=v_B\cdot v_D$, where $v_B$ and $v_D$ are the four-velocities of the $B$ and $D$~mesons, respectively. Alternatively, $q^2=(p_\ell+p_{\nu_\ell})^2$, the 4-momentum squared of the lepton neutrino system, can be used. These quantities are related by
\begin{equation}
    w=\frac{m_B^2+m_D^2-q^2}{2m_B m_D}~\label{eq:w},
\end{equation}
 where $m_B$ and $m_D$ are the $B$ and $D$ meson masses.\footnote{Natural units ($c = \hbar = 1$) are used throughout this paper.}
The range of $w$ is bounded by the zero recoil point ($w=1$), where the $D$~meson is at rest in the $B$~meson rest frame and by
\begin{equation}
      w_\mathrm{max} = \frac{m_B^2+m_D^2}{2m_Bm_D},
\end{equation}
where the entire kinetic energy is transferred to the $D$~meson. Averaging over charged and neutral modes, we find $w_{\mathrm{max}} = 1.59$.

For massless leptons, the expression of the differential decay width as a function of $w$ is~\cite{Neubert:1993mb}
\begin{equation}
    \begin{split}
      \frac{d \Gamma(B\to D\ell\nu_\ell)}{d w} = ~& \frac{G^2_\mathrm{F} m^3_D}{48\pi^3}(m_B+m_D)^2 (w^2-1)^{3/2}     \\
    &  \eta_\mathrm{EW}^2 ( 1+ \delta^{+,0}_{\textrm{C}})\mathcal{G}^2(w)|V_{cb}|^2~,
    \end{split}
    \label{eq:rate}
\end{equation}
where $G_\mathrm{F}$ is Fermi's constant, $\eta_\mathrm{EW}=1.0066 \pm 0.0050$~\cite{ ParticleDataGroup:2024cfk, Sirlin:1981ie} is the electroweak correction arising from short-distance dynamics~\cite{Sirlin:1981ie}, and $\mathcal{G}(w)$ contains the vector form factor function $f_+(w)$,
\begin{equation}
  \mathcal{G}^2(w) = \frac{4r}{(1+r)^2} f^2_+(w)~, \label{eq:ff}
\end{equation}
with $r=m_D/m_B$.
In addition to the well-understood short-distance electroweak correction, virtual photon exchanges between the $D$~meson and the lepton give rise to long-distance interactions if the final state particles carry electric charge. 
The resulting Coulomb factor is not precisely known, but it can be estimated to be $\delta^0_{\mathrm{C}} = \alpha \pi = 0.023$~\cite{PhysRevD.41.1736} for the $B^0\to D^-\ell^+\nu_\ell$ mode and $\delta^+_{\mathrm{C}} = 0$ for $B^+\to \bar D^0\ell^+\nu_\ell$, where $\alpha$ is the fine-structure constant.

In this paper, we will employ two parametrizations of $\mathcal{G}(w)$: First, we consider the parameterization by Bourrely, Caprini and Lellouch (BCL)~\cite{Bourrely:2008za} which expresses the form factors as expansions of $z$,
\begin{equation}
  z(q^2,t_0)=\frac{\sqrt{t_+-q^2}-\sqrt{t_+-t_0}}{\sqrt{t_+-q^2}+\sqrt{t_+-t_0}}.
\end{equation}
The vector and scalar form factor functions truncated at the order~$N$ are
\begin{equation}
    f_+(q^2)=\frac{1}{1-q^2/m^2_+}\sum_{k=0}^{N-1} a_k\left[z^k-(-1)^{k-N} \frac{k}{N} z^{N}\right]
    \label{eq:BCL1}
\end{equation}
and
\begin{equation}
    f_0\left(q^2\right)=\frac{1}{1-q^2 / m_0^2} \sum_{k=0}^{N-1} b_k z^k,
    \label{eq:BCL2}
\end{equation}
where $m_{+/0}$ are the masses of the closest $B_c$ resonances. The Flavor Lattice Averaging Group (FLAG)~\cite{FlavourLatticeAveragingGroupFLAG:2024oxs} presents an average of the FNAL/MILC~\cite{Lattice:2015rga} and HPQCD~\cite{Na:2015kha} lattice QCD calculations of the $B\to D$~form factors using Eqs.~\eqref{eq:BCL1} and \eqref{eq:BCL2} with $N=3$. FLAG provides an expansion for $(1 - q^2/m_{+,0}^2)f_{+,0}(q^2)$, absorbing the factors $(1 - q^2/m_{+,0}^2)$ into the parametrization. While the differential decay rate depends explicitly on the vector form factor $f_+(q^2)$ only, we include the calculation of the scalar function in our analysis as both are connected at maximum recoil $w_\mathrm{max}= 1.59$ through the kinematic constraint
\begin{equation}
  f_0(w_\mathrm{max}) = f_+(w_\mathrm{max}). \label{eq:kinematic}
\end{equation}

Alternatively, we use the expression obtained by Caprini, Lellouch, and Neubert (CLN)~\cite{Caprini:1997mu} to compare with previous experimental results~\cite{BaBar:2009zxk,Belle:2015pkj}. The CLN approach reduces the number of free parameters by adding multiple dispersive constraints based on spin- and heavy-quark symmetries,
\begin{equation}
  \mathcal{G}(y)= \mathcal{G}(1)\big(1 - 8 \rho^2 y + (51 \rho^2 - 10 )
  y^2 - (252 \rho^2 - 84 ) y^3\big),
  \label{eq:CLN}
\end{equation}
where
\begin{equation}
  y(w) = \frac{\sqrt{w + 1} - \sqrt{2} }{\sqrt{w + 1} + \sqrt{2} }.
\end{equation}
The single free parameter is $\rho^2$ while $\mathcal{G}(1)$ can be calculated by nonperturbative methods. The precision of this approximation is estimated to be better than 2\% by the original authors, which is close to the current experimental accuracy of $|V_{cb}|$.

\section{Experimental procedure} \label{sec:exp}

\subsection{Data and simulated samples}

The Belle~II detector~\cite{Abe:2010sj} operates at the SuperKEKB asymmetric-energy  electron-positron collider~\cite{superkekb}, located at KEK in Tsukuba, Japan. It consists of nested detector subsystems arranged around the beam pipe in a cylindrical geometry. The innermost subsystem is the vertex detector, which includes two layers of silicon pixel detectors and four outer layers of silicon strip detectors. In the data set used in this analysis, the second pixel layer is instrumented in only a part of the solid angle. Most of the tracking volume consists of a helium and ethane-based small-cell drift chamber (CDC). Outside the drift chamber, a Cherenkov-light imaging and time-of-propagation detector (TOP) provides charged-particle identification in the barrel region. In the forward end cap, this function is provided by a proximity-focusing, ring-imaging Cherenkov detector with an aerogel radiator (ARICH). Further out is the electromagnetic
calorimeter (ECL), consisting of a barrel and two end cap sections made of CsI(Tl) crystals. A uniform 1.5~T magnetic field is provided by a superconducting solenoid situated outside the calorimeter. Multiple layers of scintillators and resistive-plate chambers, located between the magnetic flux-return iron plates, constitute the $K^0_L$ and muon identification system (KLM).

We use the experimental data collected from 2019 to 2022 at the c.m. energy of the $\Upsilon(4S)$~resonance. 
This dataset has an integrated luminosity of \lumi~\cite{Belle-II:2019usr} and contains \NBBval $\Upsilon(4S)$ events. In addition, a 42~fb$^{-1}$ data sample 
collected at a c.m.\ energy 60 MeV below the $\Upsilon(4S)$ resonance is used to measure the continuum background.

We use samples of Monte-Carlo-simulated (MC) events equivalent to at least four times the data luminosity. The sample of $\Upsilon(4S)\to B\bar B$~events in which $B$~mesons decay generically is generated with EvtGen~\cite{Lange:2001uf}, interfaced with PYTHIA~\cite{Sjostrand:2007gs}. The branching fractions and form factors of semileptonic $B$~decays are adjusted to the most recent experimental results~\cite{ParticleDataGroup:2024cfk}. The $B\to D^*\ell\nu_\ell$ form factor is tuned to the parameters obtained in Ref.~\cite{Belle:2023bwv}. For $B\to D\ell\nu_\ell$ we use the form factor measured in Ref.~\cite{Belle:2015pkj}. We also include presently unmeasured semileptonic modes ($B\to D\eta\ell\nu_\ell$ and $B\to D^* \eta\ell\nu_\ell$), generated according to a phase space model, in equal parts to account for the discrepancy between the sum of known exclusive semileptonic branching fractions and the inclusive semileptonic $B$~decay rate. $D$~meson decays are also generated by EvtGen. Continuum background from the processes $e^+e^-\to q\bar q$ ($q = u, d, s, c$) is simulated with KKMC~\cite{Ward:2002qq} interfaced with PYTHIA~\cite{Sjostrand:2007gs}, while low-multiplicity contributions from $e^+e^-\to\tau^+\tau^-$ are generated using a combination of KKMC and TAUOLA~\cite{Jadach:1990mz}. A full detector simulation based on GEANT4~\cite{ref:geant4} is applied to the MC~events. These simulated events are overlaid with run-dependent random trigger data for accurate modeling of beam backgrounds.

\subsection{Event reconstruction}

We require charged particles (tracks) to originate from the interaction point and to have a distance of closest approach of less than 3~cm along the $z$~axis (parallel to the beams) and less than 1~cm in the transverse plane. We further require charged tracks to lie within the CDC angular acceptance ($17^\circ<\theta<150^\circ$ with $\theta$ the polar angle with respect to the $z$ axis) and to have transverse momenta greater than 50~MeV. Photons 
are reconstructed in the same angular region from in-time ECL energy depositions (clusters) unmatched to charged particle tracks. Neutral pions are searched for in their decay to two photons. All quantities are defined in the laboratory frame unless otherwise stated.

At least three charged tracks are required to suppress Bhabha scattering and other low-multiplicity processes. We select events with a ratio $R_2$ of the second to the zeroth Fox-Wolfram moment below 0.5~\cite{Fox:1978vu} to reduce backgrounds from continuum. The visible energy, i.e., the sum of energies associated to all tracks and clusters observed in the event, is required to be between 6~GeV and 10.5~GeV. These selections retain $97\%$ of the signal.

Electron candidates are identified using a boosted decision tree classifier (output $P_e > 0.9$), which combines sub-detector likelihoods and ECL observables, with the ratio of the energy measured in the ECL to the track momentum being the most discriminating variable, and are required to have c.m.\ frame momenta $p^*$ between 0.6~GeV and 2.4~GeV. We partially recover bremsstrahlung by associating colinear photons with the electron candidate. Muon candidates are identified using a likelihood-based discriminator $P_\mu = L_\mu/(L_e + L_\mu + L_\pi + L_K + L_p + L_d)$ where the likelihoods
$L_{i}$ for each charged-particle hypothesis combine particle identification information from all
detectors except the silicon trackers. We require $P_\mu > 0.9$,  and select muons in the $p^*$~range from 1.0~GeV to 2.4~GeV. The lepton candidate must be within the angular acceptance of the ECL barrel region ($32^\circ<\theta<127^\circ$). Kaons are identified using information from all subdetectors except the silicon tracker, with a requirement $P_K > 0.1$ on the likelihood-based discriminator~\cite{chargedpid}.
The identification efficiencies are 98\% for electrons, 95\% for muons, and 95\% for kaons. The rates for misidentifying pions as electrons, muons, or kaons are 0.1\%, 4\%, and 14\%, respectively.

Candidate $D$~mesons are searched for in the decay modes $D^0\to K^-\pi^+$ and $D^+\to K^- \pi^+ \pi^+$. For $D^0$~candidates, an identified $K$~candidate is combined with an oppositely charged $\pi$ and the $K^-\pi^+$~invariant mass is required to lie within the 1.85~GeV to 1.88~GeV range. For $D^+$~candidates we require $1.86<m(K^-\pi^+\pi^+)< 1.88$~GeV. In both cases the mass ranges correspond to about three times the mass resolution. In both decay modes the hadron daughters are required to have momenta larger than $0.5$~GeV. The c.m.\ frame $D$~momentum must be below 2.4~GeV to reduce background from hadronic continuum events. Candidate $B\to D\ell\nu_\ell$ decays are formed by combining an appropriately charged lepton with a $D$~candidate.
The kinematic variables of the entire $D\ell$~decay topology are subjected to a vertex fit~\cite{Belle-IIanalysissoftwareGroup:2019dlq} and combinations with a $\chi^2$~probability below 0.05 are rejected.

Background from $B\to D^*\ell\nu_\ell$~decays (``feed-down'') is vetoed by combining the $D$~candidate with a pion and reconstructing the modes $D^{*0}\to D^0\pi^0$, $D^{*+}\to D^0\pi^+$ and  $D^{*+} \to D^+\pi^0$. Events are rejected if at least one $D^*$~candidate with $\Delta m=m_{D^*}-m_D$ in the appropriate range is found. These intervals are $[140~\mathrm{MeV},150~\mathrm{MeV}]$, $[135~\mathrm{MeV},150~\mathrm{MeV}]$ and $[135~\mathrm{MeV},145~\mathrm{MeV}]$ for the $D^{*+}\to D^0\pi^+$, $D^{*0}\to D^0\pi^0$ and $D^{*+}\to D^+\pi^0$~modes, respectively.

To further reduce contamination from continuum and $B\bar B$ backgrounds, we apply
a set of selections on seven variables. These include the modified Fox-Wolfram moment $H^{so}_{20}$~\cite{Bevan:2014iga}, $m_{Y}$, the missing momentum in the c.m.\ frame $p^*_\mathrm{miss}$, the invariant mass of the rest-of-event (ROE) system, the ROE momentum $p_\mathrm{ROE}$, the cosine of the angle between the direction of the lepton in the virtual $W$ rest frame and the direction of the $W$ boson in the $B$ rest frame $\cos\theta_{\ell,W}$, and the angle between the $D$ and $\ell$~directions in the laboratory frame $\theta_{D,\ell}$. The ROE is defined as the collection of all particles in the event that are not used in the reconstruction of the $Y=D\ell$ system. To reconstruct $\cos\theta_{\ell,W}$, the momentum vector of the $B$ meson is approximated utilizing the algorithm outlined in the reconstruction of $w$ in Sec.~\ref{sec:w}.

The selection criteria are optimized using an algorithm based on simulated annealing~\cite{annealing} to maximize the geometric mean of the expected signal significances across bins of the hadronic recoil $w$. They are determined separately for the charged and neutral modes.

\subsection{\texorpdfstring{Reconstruction of $w$}{Reconstruction of w}} \label{sec:w}

To access the variable $w$ (Eq.~\eqref{eq:w}), we use an improved version of the method described in Ref.~\cite{BaBar:2006taf}, which was previously employed in Ref.~\cite{PhysRevD.108.092013}. The $B$~meson 3-momentum lies on a cone around the $Y$ direction defined by
\begin{equation}
  \cos\theta_{BY} = {2\, E_\mathrm{Beam} E_Y - m_B^2 - m_Y^2 \over 2|\vec p_B||\vec p_Y|}~, \label{eq:costheta}
\end{equation}
where $E_Y$, $|\vec p_Y|$, and $m_Y$ are the reconstructed energy, momentum, and invariant mass of $Y$, respectively, 
$E_\mathrm{Beam}$ is the beam energy in the c.m.\ frame, and $|\vec p_B|$ is the $B$ meson momentum, computed from $m_B$ and $E_\mathrm{Beam}$.
Next, we consider several configurations of the $B$~momentum on this cone. For each potential $\hat p_B$ direction, the kinematic variable $w = v_B \cdot v_D$ is calculated and our estimate of $w$ is the weighted average over these configurations. The weights are taken to be
\begin{equation}
    u=\frac{1}{2}(1-\hat p^*_B\cdot\hat p^*_\mathrm{ROE})\sin^2\theta^*_B~,
\end{equation}
where $\hat p^*_\mathrm{ROE}$ is the unit vector of the c.m.\ frame rest-of-event momentum, and $\theta^*_B$ is the angle of the $B$~meson direction with respect to the beam axis in the c.m.\ frame.

The weight assigned to the different possible $B$~directions accounts for the expectation that the $B$~direction follows the direction opposite to that of the ROE~system momentum. We also take into account that the $B$~direction is expected to follow a $\sin^2\theta^*_B$ distribution since the $\Upsilon(4S)$ is produced in $e^+e^-$ annihilation. The $w$ resolution resulting from this improved kinematic calculation is estimated to be 0.038 (0.034) for the charged (neutral) mode.

We split the reconstructed $B\to D\ell\nu_\ell$ sample into ten equal-width bins in $w$ with bin boundaries at $w=1,1.06,1.12,1.18,1.24,1.3,1.36,1.42,1.48,1.54, ~\mathrm{and}~w_\mathrm{max}$.

\subsection{Corrections to simulated data}
\label{sec:corrections}
To validate the description of signal and background processes in simulation, we define and reconstruct a set of control samples to assess the agreement between experimental and simulated data and to identify potential selection biases.

For the signal validation, we use hadronic $B \to D \pi$ decays, taking the $\pi$ as a substitute for the lepton in $B\to D\ell\nu_\ell$ decays. Effects from missing energy, missing momentum, and the ROE-system are studied using partially reconstructed $B \to D \pi$ decays that originate from $B \to D^*(\to D \pi) \pi$ decays, where the unreconstructed slow pion introduces missing energy. Analogously, we partially reconstruct $B \to K^* \ell$ within a sample of fully reconstructed $B \to K^*(\to K \pi)J/\psi(\to \ell \ell)$ decays. The unreconstructed second lepton introduces missing energy and momentum, allowing validation of the ROE and the kinematic variable reconstruction.

We validate all selections by comparing selection efficiencies in data and MC, finding good agreement for all but the vertex fit $\chi^2$-probability, for which we observe a mismodeling that is present in all signal validation samples. We derive data-driven correction factors from the $B \to K^*\ell$ control sample, by using the ratio of expected to observed candidates. The correction factors are binned in the vertex fit $\chi^2$ probability, and independently validated by applying them to the $B\to D \pi$ control sample, in which the ratio of signal selection efficiencies in data over MC is corrected from $\varepsilon_\mathrm{data} / \varepsilon_\mathrm{MC} = 1.010\pm 0.002$ to $\varepsilon_\mathrm{data} / \varepsilon_\mathrm{MC} = 0.999 \pm 0.002$. After validation, we apply the correction factors to the signal component in the $B\to D\ell\nu_\ell$ sample.

To validate background contributions, we define background-enriched control samples for each of the four major background categories. Continuum events are isolated using off-resonance data, providing a sample without $B$ meson decays. Misreconstructed $D$ meson backgrounds are studied by selecting a sideband in $m(K\pi (\pi))$, which yields a sample dominated by incorrectly reconstructed $D$ candidates. A sample enriched with $B \to D^* \ell \nu_\ell$ backgrounds is obtained by inverting the veto requirement designed to suppress these events. Finally, a ``same-charge'' control sample is formed from pairs with intentionally incorrect charge combinations, such as $D^+ \ell^+$,  to provide a sample for studying backgrounds with correctly reconstructed $D$ mesons that do not fall into any of the previous categories.

Observed biases in the comparison between simulation and data in the reconstructed energy and mass of the $D\ell$ system extend to the $\cos\theta_{BY}$ distribution.  To correct for these biases, we extract calibration factors using the distributions of $E^*_Y$ and $m_Y$ in the background-enriched control samples defined above.  The calibration factors are computed independently for each of the true $D$, fake $D$, continuum and $B \rightarrow D^* \ell \nu_\ell$ control samples, from their respective background control sample.
The correction factors are most significant for the true $D$ component, for which we apply an average weight of 0.96 with a standard deviation of 0.10, while the averages of other corrections remain between 0.999 and 1.003. Applying the correction factors to the signal reconstruction significantly improves the agreement between data and simulation in the $m_Y$, $E^*_Y$, and $\cos\theta_{BY}$ distributions.

Additional corrections are applied to account for mismodeling in the simulation of particle identification efficiencies and misidentification likelihoods. These corrections are evaluated by comparing data and simulation using pure samples of electrons and muons from $J/\psi \to \ell^+ \ell^-$ and low-multiplicity $e^+ e^- \to \ell^+ \ell^- (\gamma)$ and $e^+ e^- \to e^+ e^- \ell^+ \ell^-$ processes.
To calibrate charged kaon identification, we use a sample of kaons from $D^{*+} \to D^0(\to K^- \pi^+) \pi^+$ decays and misidentified pions from $K^0_S \to\pi^+ \pi^-$. The correction factors are binned in the laboratory frame momenta and polar angles of the particle tracks.

\begin{figure*}
    \centering
    \includegraphics[width=.7\columnwidth]{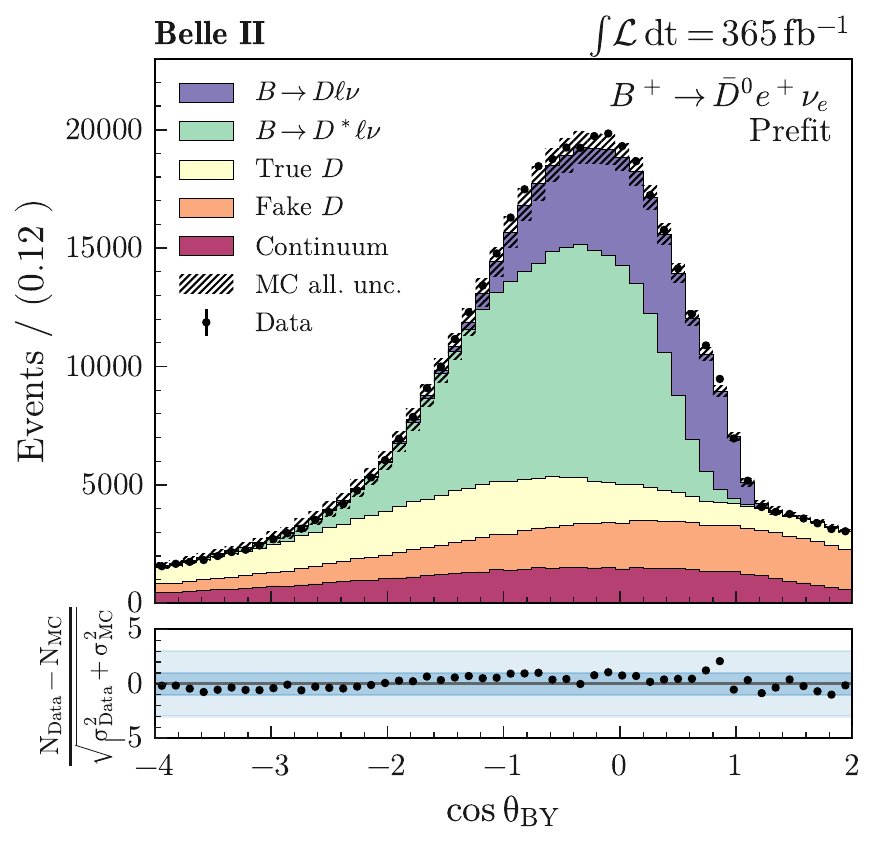}
    \includegraphics[width=.7\columnwidth]{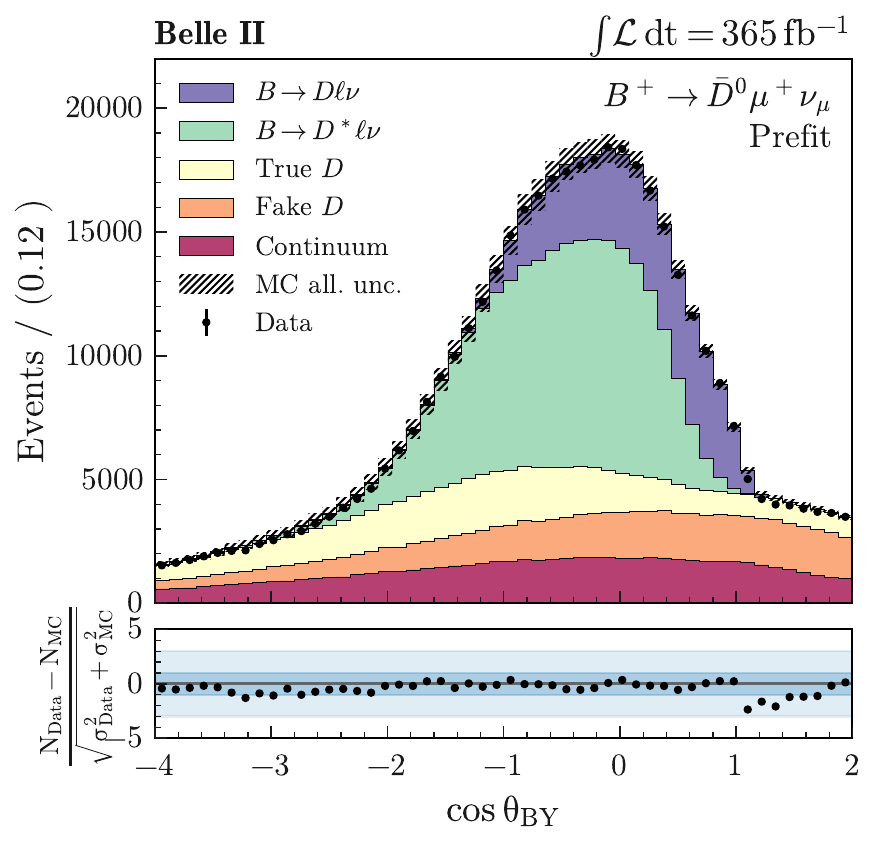}\\
    \includegraphics[width=.7\columnwidth]{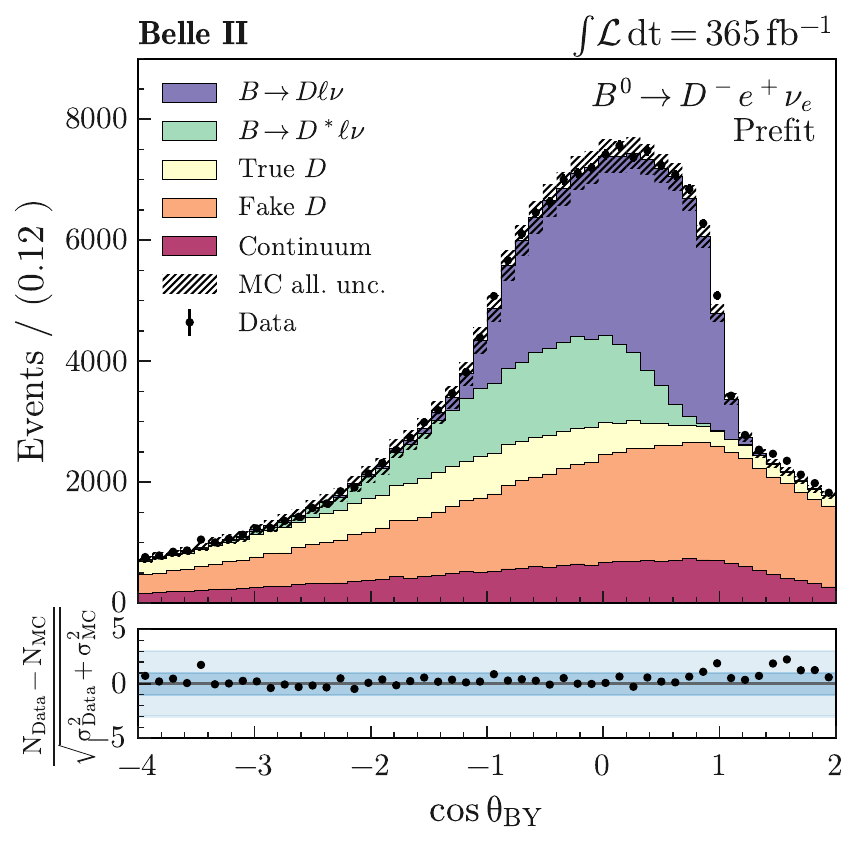}
    \includegraphics[width=.7\columnwidth]{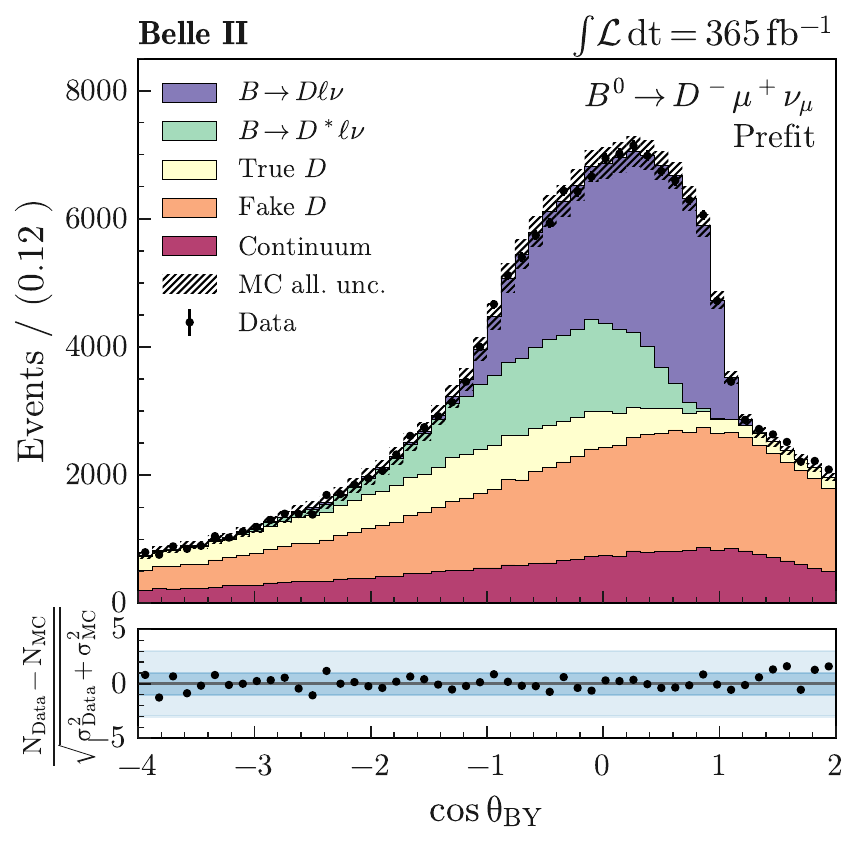}
    \caption{Distribution of $\cos\theta_{BY}$ after signal selection shown separately for the charged and neutral modes and electron and muon events. The Belle II data are the points with error bars. The stacked histograms show the expected distributions prior to the fit, normalized to the luminosity and after applying the corrections described in Sec.~\ref{sec:corrections}. Section~\ref{sec:fit} describes the different components. The hatched area represents the statistical and systematic uncertainties of the simulated samples. The panels at the bottom of each distribution show the difference between data and simulation divided by the combined uncertainty.} \label{fig:prefit}
\end{figure*}

Figure~\ref{fig:prefit} shows the result of the signal selection summed over bins of $w$ after the corrections have been applied. Additional validation plots for kinematic variables in control samples are provided in the Supplemental
Material~\cite{supplemental}.

\subsection{Signal extraction} \label{sec:fit}

The signal yield is extracted using a maximum likelihood fit to $\cos\theta_{BY}$ (Eq.~\eqref{eq:costheta}) after applying all selections described above. For correctly reconstructed signal decays and assuming perfect resolution, we expect $\cos\theta_{BY}$ to lie between $-1$ and $1$.
Due to detector resolution effects, signal events can be slightly smeared beyond the physical range. For background events, $\cos\theta_{BY}$ can take unphysical values—particularly when the assumption that only a single neutrino is missing is not valid, such as in candidates from $B \to D^{**} \ell \nu_\ell$ decays.

We use 10 equal-width bins of $\cos\theta_{BY}$ in the range $[-4,2]$ as the fitting region to separate signal and background events. If the lowest bin in $\cos\theta_{BY}$ is empty, it is iteratively merged with the next-lowest bin until all bins contain events. The choice of 10 initial equal-width bins is sufficient to capture the key features of the distributions, and using more bins does not lead to further improvements in the fit precision.

In the fit, we consider five components: $B\to D\ell\nu_\ell$~signal, feed-down background from the decay $B\to D^*\ell\nu_\ell$, combinations with a correctly reconstructed $D$~meson that does not originate from a $B\to D^{(*)}\ell\nu_\ell$~decay (``true $D$"), and combinatorial background in which the $D$~meson is misreconstructed (``fake $D$"). Finally, candidates from non-$B \bar B$ events from processes such as $e^+ e^-\to q\bar q$ ($q=u,d,s,c$) and $e^+e^-\to\tau^+\tau^-$ are combined into the continuum background category. The fit templates for all components are derived from simulation, with systematic uncertainties incorporated through nuisance parameters that modify both the template shape and normalization. The nuisance parameters are varied in the fit together with the signal and background yields~\cite{Heinrich:2021gyp}.

The fit is performed simultaneously in ten bins of $w$ using the bin boundaries defined in Sec.~\ref{sec:w}, in the electron and the muon samples, and for the charged and the neutral modes, to determine the differential decay widths $\Delta\Gamma_i/\Delta w$ of the $D^0e$, $D^0\mu$, $D^+e$ and $D^+\mu$ subsamples. The effects of bin-to-bin migration due to $w$ resolution, which can reach the level of $20\%$ between neighboring bins, are accounted for by constructing signal templates based on their generated values of $w$.

The multiplicative nature of efficiency corrections and global normalization factors can cause a bias when averaging across bins~\cite{DAgostini:1993arp}. To mitigate this effect, the average across modes is obtained directly from the simultaneous fit by linking the signal normalization parameters across channels. For the determination of branching fractions in individual modes and for tests of lepton flavor universality, the fit is performed with separate signal normalization parameters for each subsample.


The fit determines the differential decay width $\Delta\Gamma_i/\Delta w$ in the $i^\mathrm{th}$ bin of $w$,
\begin{equation}
\label{eq:fitdgdw}
 \frac{\Delta\Gamma_i}{\Delta w}=\frac{N_i}{N_B\mathcal{B}(D)\epsilon_i\tau_B (1+ \delta^{0,+}_C)\Delta w}~,
 \quad i=1,\dots,10~
\end{equation}
which are treated as free parameters in the fit.
Here, $B$ stands for either $B^+$ or $B^0$ and $N_B$ is the number of $B$~mesons in the experimental data sample,
\begin{eqnarray}
  N(B^+) & = & 2f^{+-}N_{\Upsilon(4S)}~, \\
  N(B^0) & = & 2f^{00}N_{\Upsilon(4S)}~,
\end{eqnarray}
with $N_{\Upsilon(4S)}$ the number of $\Upsilon(4S)$~events. $f^{+-}$ ($f^{00}$) is the fraction of $\Upsilon(4S)$ decays to $B^+B^-$ ($B^0\bar B^0$). $N_i$ is the yield in the $i^\mathrm{th}$~bin of generated $w$ before detector effects, $\mathcal{B}(D)$ is the subdecay branching fraction (either $D^0\to K^-\pi^+$ or $D^+\to K^-\pi^+\pi^+$), $\epsilon_i$ is the overall efficiency after applying all selections in the given bin and sub-sample, $\tau_B$ is either the $B^+$ or the $B^0$~lifetime, $\delta^{0,+}_C$ is the long-distance QED correction introduced in Eq.~\eqref{eq:rate},
and $\Delta w$ is the bin width.

\subsection{Systematic uncertainties} \label{sec:syst}

The uncertainties in the nuisance parameters are used to determine the corresponding systematic uncertainties and their correlations.

\subsubsection{\texorpdfstring{$B$ lifetime}{B lifetime}}

The $B^0$ and $B^+$~lifetimes are taken to be $1.517\pm 0.004$~ps and $1.638\pm 0.004$~ps~\cite{ParticleDataGroup:2024cfk}, respectively, and included in the likelihood by two nuisance parameters.

\subsubsection{\texorpdfstring{$B\to D^{(*)}\ell\nu_\ell$ form factors}{B to D(*) l nu form factors}}

Signal templates are divided by the generator value of $w$, and variations in the form factor in simulation affect the measured rates in two ways: (1) the $\cos\theta_{BY}$ shape of the signal template in each reconstructed bin of $w$ can vary as a function of the input form factor. This affects both $B \rightarrow D$ and $B \rightarrow D^*$ templates.
(2) The signal templates contain tails that migrate into neighboring bins of $w$, and therefore, varying the input distribution affects the magnitude of the bin-to-bin migrations.

To propagate these effects into the fit, the covariance matrices of the $B\rightarrow D$ and $B \rightarrow D^*$ form factors in Refs.~\cite{Belle:2023bwv,Belle:2015pkj} are decomposed into their eigendirections. A nuisance parameter is assigned to every eigencomponent and constrained to vary within the corresponding uncertainty. Variations of the $B\to D\ell\nu_\ell$~form factor are fully correlated over all signal templates, while variations of the $B\to D^*\ell\nu_\ell$~parameters simultaneously affect all feed-down background templates. There are eight (five) parameters for the $D\ell\nu_\ell$ ($D^*\ell\nu_\ell$) form factor shapes.

\subsubsection{\texorpdfstring{$X_c \ell \nu_\ell$ model}{X_c l nu model}}
Exclusive semileptonic $B$~decay modes other than $B\to D^{(*)}\ell\nu_\ell$ are adjusted to their measured branching fractions~\cite{ParticleDataGroup:2024cfk} and allowed to vary within their experimental uncertainty. Unmeasured modes ($B\to D\eta\ell\nu_\ell$ and $B\to D^{*}\eta\ell\nu_\ell$) are added to fill the gap to the inclusive semileptonic $B$~decay rate in equal parts and are allowed to vary in the fit with a $100\%$ uncertainty. In total, 24 nuisance parameters are used to account for systematic variations of the $X_c\ell\nu_\ell$~model.

\subsubsection{\texorpdfstring{$D$~meson branching fractions}{D meson branching fractions}}
The $D^0\to K^-\pi^+$ and the $D^+\to K^-\pi^+\pi^+$ branching fractions are rescaled to the latest world average values and allowed to vary within their respective uncertainties~\cite{ParticleDataGroup:2024cfk} (two parameters).

\subsubsection{Charged particle tracking}

Differences between track finding efficiencies in data and simulation for tracks above 300~MeV are determined in $e^+ e^- \rightarrow \tau^+ \tau^-$ events, in which one $\tau$ lepton decays into three charged tracks.
The resulting uncertainty is 0.24\% for each charged track. Adding this error linearly results in a total tracking uncertainty of 0.72\% (0.96\%) for the charged (neutral) mode. This uncertainty is considered to be fully correlated between all templates and modeled by a single nuisance parameter.

\subsubsection{\texorpdfstring{$N_{\Upsilon(4S)}$ and $f^{+-}/f^{00}$}{N Upsilon(4S) and f+- / f00}}
The integrated luminosities of the on- and off-resonance data samples in this analysis are determined by counting Bhabha events and using the well-known cross section of this process. In the next step, the number of $\Upsilon(4S)$~decays in the sample of selected hadronic events is estimated by subtracting the scaled off-resonance data, and \NBB is found. We include the uncertainty on $N_{\Upsilon(4S)}$ as a nuisance parameter, fully correlated between final states.\\
In addition to this, we use two nuisance parameters affecting the $B^0/B^\pm$ fractions: one for the ratio $f^{+-}/f^{00}$, constrained to $1.052 \pm 0.031$~\cite{HeavyFlavorAveragingGroupHFLAV:2024ctg}, and another for the fraction of non-$B\bar{B}$ decays of the $\Upsilon(4S)$, $f_{\slashed{B}} = 0.0027^{+0.0138}_{-0.0002}$~\cite{HeavyFlavorAveragingGroupHFLAV:2024ctg}, where $f_{+-} + f_{00} + f_{\slashed{B}} = 1$.

\subsubsection{\texorpdfstring{Background $w$ modelling}{Background w modelling}}
The $w$ distributions of the feed-down $B\to D^*\ell\nu_\ell$, true $D$, fake $D$, and continuum backgrounds are validated using dedicated control samples: the inverted $D^*$ veto sample, $D^+\ell^+$ wrong-sign combinations, the $m_D$ sidebands, and off-resonance data. These samples allow us to assess the accuracy of the MC templates and assign appropriate uncertainties.

For the lowest $w$ bin, we conservatively assign a 30\% uncertainty for both charged and neutral $B$ modes, as this bin is poorly populated and the estimates of the uncertainty fluctuate. For the remaining $w$ bins, uncertainties are derived from binned likelihood fits performed separately on each control sample and $B$ mode, where shape uncertainties capture discrepancies between experimental and Monte Carlo data.

We obtain constraints on the bin-by-bin normalizations by performing a fit to the $w$ distribution in each control sample. For the fake $D$, feed-down, and continuum background components, we obtain bin-by-bin variations constrained to 3.2\% (2.0\%), 2.4\% (10.6\%), and 2.6\% (2.9\%) in the charged (neutral) mode, respectively. In addition to these Gaussian-constrained bin-by-bin variations, each background component also includes an unconstrained overall normalization parameter shared across all $w$ bins.

We observe good agreement between the different lepton flavor modes in the control samples and, therefore, use joint bin-by-bin normalizations, leading to 20 nuisance parameters per sample (continuum, fake $D$, $B \rightarrow D^* \ell \nu_\ell$) i.e. 60 in total. Due to data-MC disagreements in the wrong-charge control sample, we leave the bin-by-bin variations for the true $D$ component unconstrained and split between modes.

\subsubsection{\texorpdfstring{$(E^*_Y,m_Y)$ background reweighting}{(Ey, mY) background reweighting}}

To estimate the systematic uncertainty associated with the $(E^*_Y, m_Y)$ background reweighting correction, we resample the correction factors derived from control samples. We generate toy sets of correction factors by applying uncorrelated Gaussian smearing terms based on the statistical uncertainties of the control samples, along with correlated Gaussian smearing terms with widths equal to the full correction magnitude.

These variations are then propagated to the signal region bins, and their effects, along with correlations, are incorporated into a covariance matrix. This
matrix is subsequently factorized into a canonical form using eigendecomposition and represented using nuisance parameters. We retain the first 30 eigenvectors as correlated shape uncertainties. Including additional terms does not change the error estimate.

\subsubsection{Particle identification}
The efficiencies and misidentification rates of the electron, muon and kaon identification requirements are determined from various data in bins of laboratory frame momentum and polar angle. As the systematic correlations between these bins are unknown, different assumptions are tested. Among the different assumptions, full correlation results in the largest systematic uncertainty and is therefore chosen.

The corresponding modifiers are obtained in a similar manner as for the $\cos\theta_{BY}$ correction: the systematic variations are then propagated to the signal region, and the resulting covariance matrix is decomposed into eigenvectors. The leading five eigenvectors are then used as nuisance parameters to represent correlated shape uncertainties.

\subsubsection{\texorpdfstring{Vertex fit $\chi^2$ correction}{Vertex fit chi2 correction}}
The systematic error related to the vertex fit $\chi^2$~probability correction factors, derived from the $B^0\to K^* \ell$~control sample, is modeled as a fully correlated normalization uncertainty on the signal templates. The width of the corresponding nuisance parameter is determined by varying the control sample within its statistical uncertainties, recalculating the correction factors, and propagating the resulting changes to the observed signal efficiency of the selection.

\subsubsection{Simulation sample size}
The templates are allowed to vary within their statistical uncertainty to account for the limited sample size of the simulation. For the nominal fit with ten bins in $\cos\theta_{BY}$, ten bins in $w$ and four subsamples, this corresponds to 380 nuisance parameters after combining bins to avoid empty bins. The per-bin scale factors are uncorrelated.

\subsubsection{Electroweak corrections}

As discussed in Sec.~\ref{sec:theo}, the charged current vertex in the decay $B\to D\ell\nu_\ell$ can receive both short- and long-distance electroweak corrections that affect the extraction of $|V_{cb}|$. The modifier on the short-distance electroweak correction $\eta_\mathrm{EW}$ is fully correlated between charged and neutral $B$~decays. The long-distance Coulomb correction resulting from virtual photon exchange between the $D$ meson and the lepton affects only the neutral $B$~mode. Given that the value of this correction is not precisely known, we assume $\delta^{0}_C = \delta^{+}_C = 0$ and assign a nuisance parameter with width of 0.023 to $\delta^{0}_C$ corresponding to the estimated order of magnitude of such effects~\cite{PhysRevD.41.1736} when performing the isospin average. The averaged fit yields $\delta^0_C = 0.023 \times (-0.09 \pm 0.94)$.\\[1ex]

After pruning nuisance parameters with negligible impact on the fit, the model includes 550 constrained parameters and 62 unconstrained parameters, including the (linked) signal normalization parameters across channels, bin-by-bin variations for the true $D$ background, and overall normalizations for the major background components.

\section{Results} \label{sec:res}

\begin{figure*}
    \centering
    \includegraphics[width=1.5\columnwidth]{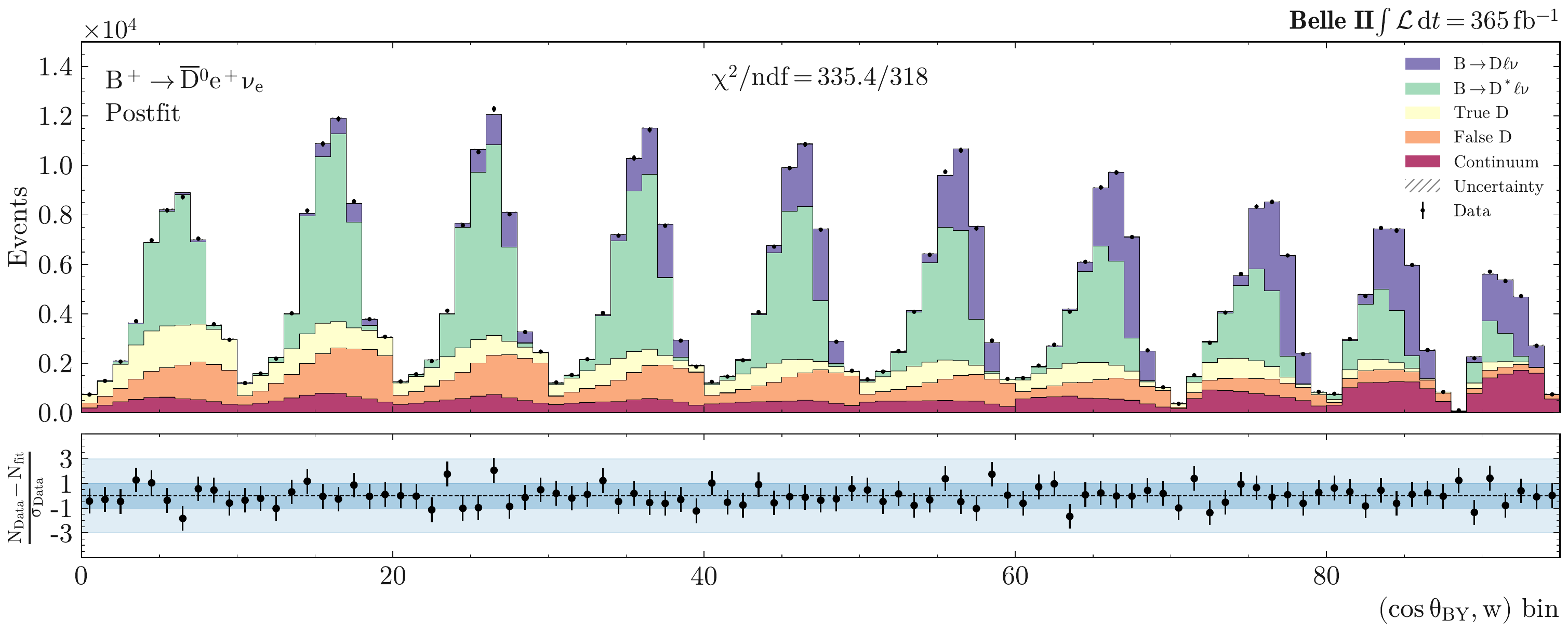}
    \includegraphics[width=1.5\columnwidth]{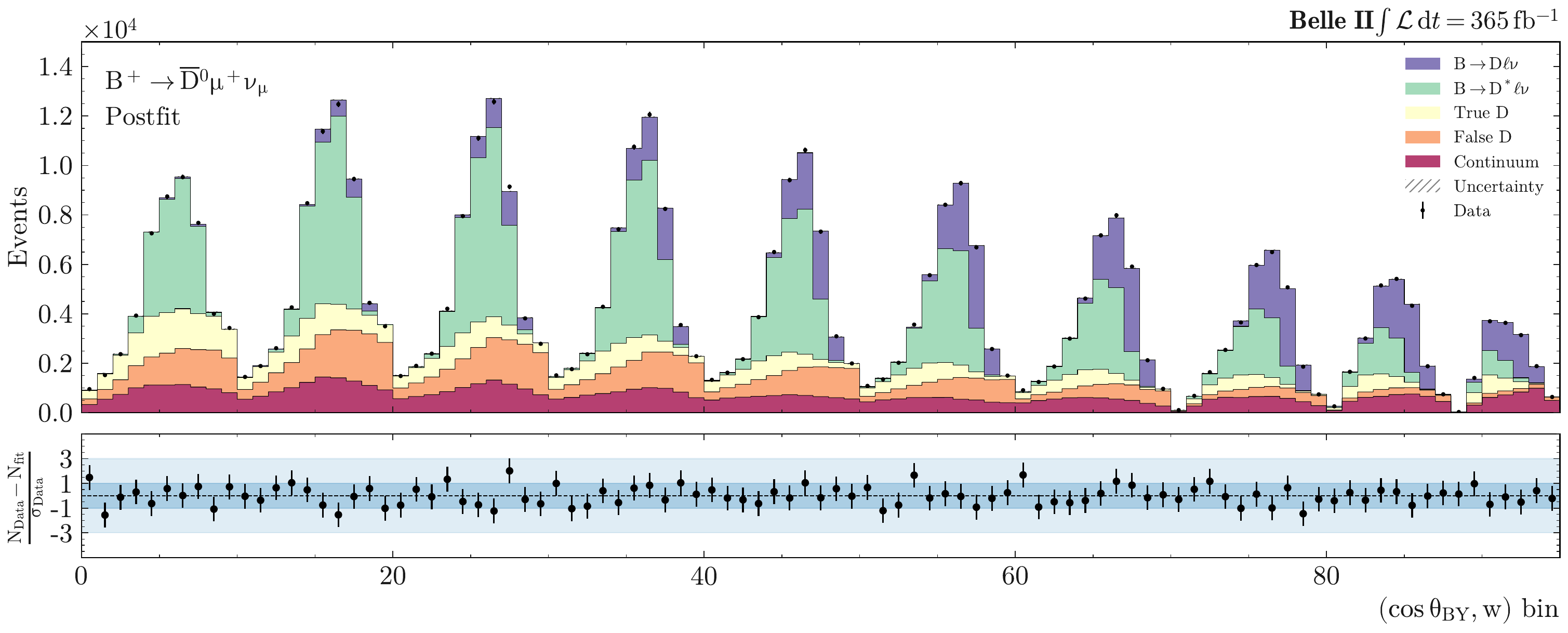}
    \includegraphics[width=1.5\columnwidth]{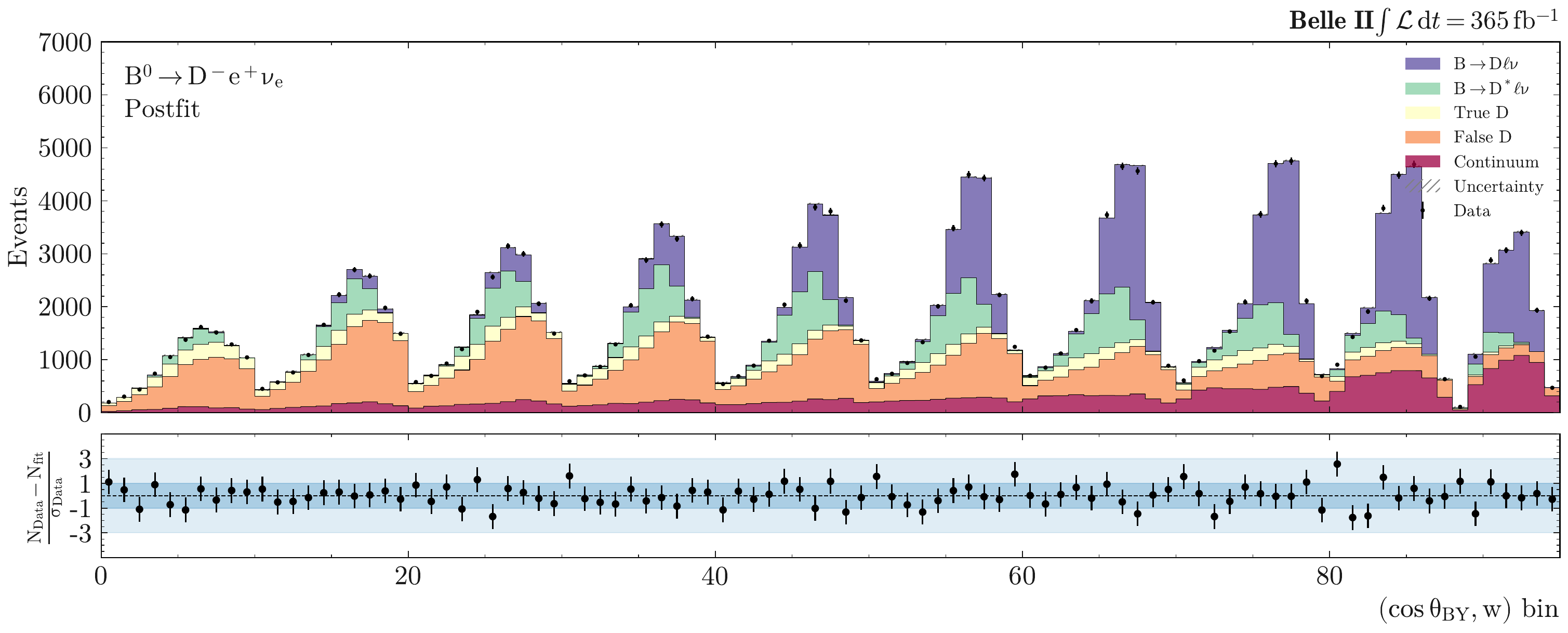}
    \includegraphics[width=1.5\columnwidth]{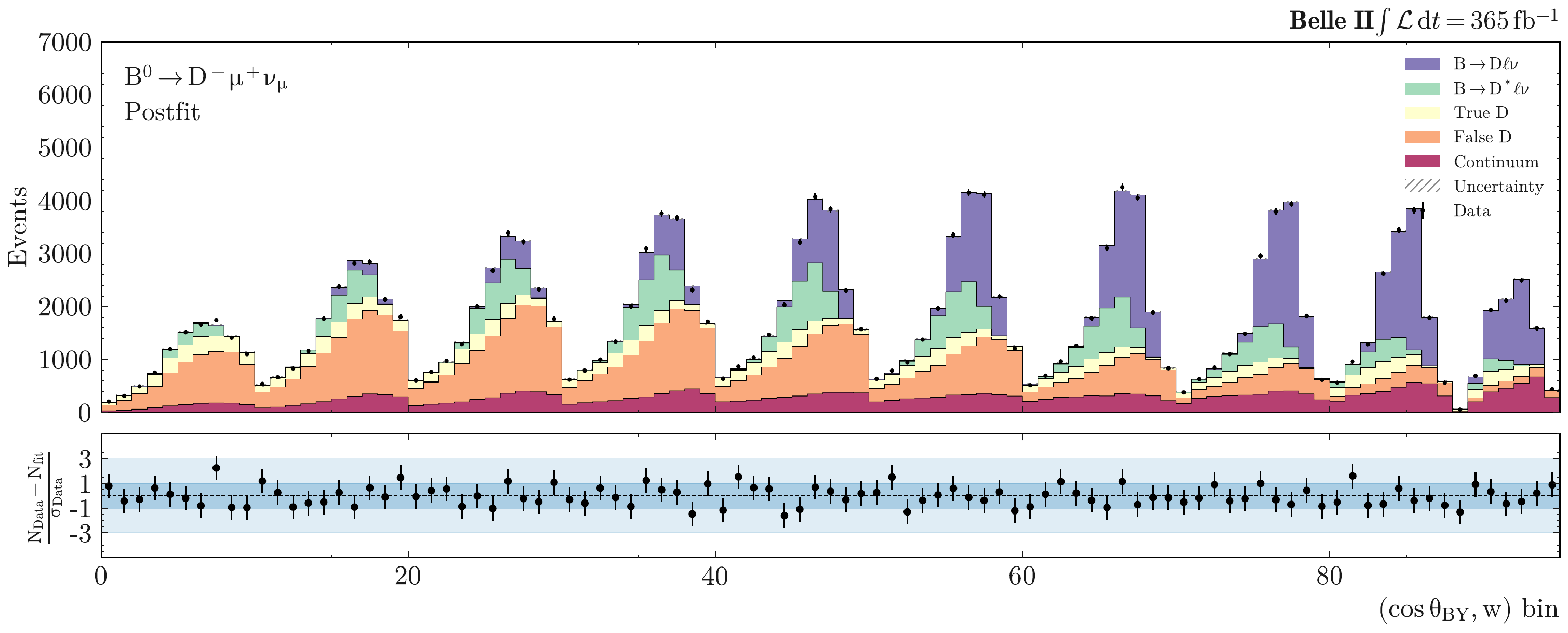}
    \caption{Fitted distributions of $\cos\theta_{BY}$ in bins of $w$ in the $D^0e$, $D^0\mu$, $D^+e$ and $D^+\mu$ samples. The two-dimensional distribution in $(\cos\theta_{BY}, w)$ is ``unrolled'', such that each group of bins corresponds to $\cos\theta_{BY}$ in a given $w$ interval, with successive groups representing increasing $w$ bins. The Belle II data are the points with error bars. The stacked histograms are simulated events normalized according to the fit result. Sec.~\ref{sec:fit} describes the different components. The panels at the bottom of each distribution show the difference between data and fit results, divided by the statistical uncertainty.} \label{fig:postfit}
\end{figure*}
\begin{table}
\caption{Fitted signal yields and semileptonic $B\to D$ branching fractions in the subsamples. The first uncertainty is statistical, and the second is systematic. The $B^+\to \bar D^0\ell^+\nu_\ell$ and $B^0\to D^- \ell^+ \nu_\ell$ branching fractions are the averages over lepton flavors. The final row represents an average over all four samples, assuming the neutral $B$ meson lifetime for all $B$ mesons.} \label{tab:BR}
\vspace{0.3cm}

\begin{tabular}{lcc}
\toprule \toprule
 & Signal yield & $\mathcal{B}~[\%]$ \\
\midrule
$B^+ \rightarrow \bar{D}^0 e^+ \nu_e$ & $75\,186$ & $2.34 \pm 0.05 \pm 0.10$ \\

$B^+ \rightarrow \bar{D}^0 \mu^+ \nu_\mu$ & $61\,259$ & $2.27 \pm 0.05 \pm 0.09$ \\

$B^0 \rightarrow D^- e^+ \nu_e$ & $47\,617$ & $2.07 \pm 0.06 \pm 0.10$ \\
$B^0 \rightarrow D^- \mu^+ \nu_\mu$ & $39\,648$ & $2.05 \pm 0.06 \pm 0.11$ \\
[1.5ex]

$B^+ \rightarrow \bar{D}^0 \ell^+ \nu_\ell$ &  & $2.31 \pm 0.04 \pm 0.09$ \\
$B^0 \rightarrow D^- \ell^+ \nu_\ell$ &  & $2.06 \pm 0.05 \pm 0.10$ \\[1.5ex]

$B \rightarrow D \ell \nu$ &  & $2.10 \pm 0.04 \pm 0.06$ \\
\bottomrule \bottomrule

\end{tabular}

\end{table}

\begin{table*}
\caption{Values of $\Delta\Gamma_i/\Delta w$ with their total uncertainty and correlation coefficients (including statistical and systematic contributions) in different bins of $w$ after combination of the four subsamples. The columns are (from left to right) the bin number, the lower and the upper edge of the $i^\mathrm{th}$~bin, the value of
$\Delta\Gamma_i/\Delta w$ in this bin with the total uncertainty, and the correlation matrix.
The value of $w_\mathrm{max} = 1.591$ is the average of the values for charged and neutral $B$ mesons.}
\label{tab:dgdwResults}

\vspace{0.3cm}
\begin{tabular}{ccc >{\centering\arraybackslash}p{4cm} p{0.8cm} p{0.8cm} p{0.8cm} p{0.8cm} p{0.8cm} p{0.8cm} p{0.8cm} p{0.8cm} p{0.8cm} p{0.8cm}}
\toprule \toprule
 & & & & \multicolumn{10}{c}{$\rho_{ij}$} \\
$i$ & $w_{i,\mathrm{min}}$ & $w_{i,\mathrm{max}}$ & \dGidw ~[$10^{-15}$GeV]   & 1 & 2 & 3 & 4 & 5 & 6 & 7 & 8 & 9 & 10 \\
\midrule
1 & 1.00 & 1.06 & $0.22 \pm 0.59$ & 1.00 & $-0.06$ & 0.15 & 0.08 & 0.07 & 0.04 & 0.03 & 0.02 & 0.00 & 0.00 \\
2 & 1.06 & 1.12 & $3.54 \pm 0.56$ &  & $\phantom{-}1.00$ & 0.13 & 0.33 & 0.26 & 0.24 & 0.22 & 0.19 & 0.13 & 0.07 \\
3 & 1.12 & 1.18 & $6.46 \pm 0.61$ &  &  & 1.00 & 0.25 & 0.44 & 0.37 & 0.36 & 0.31 & 0.22 & 0.13 \\
4 & 1.18 & 1.24 & $10.17 \pm 0.68$ &  &  &  & 1.00 & 0.37 & 0.59 & 0.52 & 0.48 & 0.34 & 0.20 \\
5 & 1.24 & 1.30 & $14.27 \pm 0.72$ &  &  &  &  & 1.00 & 0.49 & 0.67 & 0.55 & 0.41 & 0.23 \\
6 & 1.30 & 1.36 & $18.68 \pm 0.84$ &  &  &  &  &  & 1.00 & 0.58 & 0.71 & 0.49 & 0.30 \\
7 & 1.36 & 1.42 & $21.41 \pm 0.89$ &  &  &  &  &  &  & 1.00 & 0.59 & 0.60 & 0.36 \\
8 & 1.42 & 1.48 & $25.42 \pm 0.96$ &  &  &  &  &  &  &  & 1.00 & 0.49 & 0.48 \\
9 & 1.48 & 1.54 & $28.11 \pm 1.09$ &  &  &  &  &  &  &  &  & 1.00 & 0.61 \\
10 & 1.54 & $w_{\mathrm{max}}$ & $29.44 \pm 1.41$ &  &  &  &  &  &  &  &  &  & 1.00 \\
\bottomrule \bottomrule
\end{tabular}
\label{tab:dGdw}
\end{table*}

\begin{table*}
\caption{Values of $\Delta\Gamma_i/\Delta w$ with their total uncertainty in the four subsamples. $i$, $w_{i,\mathrm{min}}$ and $w_{i,\mathrm{max}}$ are the $w$-bin number, lower and upper edge of the bin, respectively. The value of
$w_\mathrm{max}$ is 1.592 for the subsamples with a charged $B$ meson and 1.589 for the subsamples with a neutral $B$ meson.  The rate is corrected for long-distance Coulomb interaction effects for the neutral mode with $\delta^0_C = 0.023 \times (-0.09 \pm 0.94)$ as determined in the fit. The correlations between the $\Delta\Gamma_i/\Delta w$~bins and samples are available on HEPData~\cite{hepdata.153613}.}
\vspace{0.3cm}
\label{tab:dGdwsubsamples}

\begin{tabular}{p{1cm} p{1cm} p{1cm} >{\centering\arraybackslash}p{2.5cm} >{\centering\arraybackslash}p{2.5cm} >{\centering\arraybackslash}p{2.5cm} >{\centering\arraybackslash}p{2.5cm}}
\toprule \toprule
 & & & \multicolumn{4}{c}{ $\dGidw~[10^{-15}\text{GeV}]$  } \\
$i$  & $w_{i,\mathrm{min}}$ & $w_{i,\mathrm{max}}$  & $B^{0} \to D^- e^+ \nu_e$ & $B^{0} \to D^- \mu^+ \nu_\mu$ & $B^+ \to \bar{D}^{0}e^+\nu_e$ & $B^+ \to \bar{D}^{0}\mu^+\nu_\mu$\\
\midrule
1 & 1.00 & 1.06 & $-0.1 \pm 0.7$ & $0.6 \pm 0.7$ & $-0.5 \pm 1.5$ & $1.8 \pm 1.7$ \\
2 & 1.06 & 1.12 & $\phantom{-}3.7 \pm 0.8$ & $3.2 \pm 0.8$ & $\phantom{-}4.1 \pm 1.3$ & $3.7 \pm 1.4$ \\
3 & 1.12 & 1.18 & $\phantom{-}6.2 \pm 0.8$ & $7.2 \pm 0.9$ & $\phantom{-}5.5 \pm 1.3$ & $6.1 \pm 1.2$ \\
4 & 1.18 & 1.24 & $\phantom{-}9.8 \pm 0.9$ & $11.0 \pm 0.9$ & $\phantom{-}8.9 \pm 1.3$ & $10.3 \pm 1.4$ \\
5 & 1.24 & 1.30 & $13.9 \pm 1.0$ & $14.8 \pm 1.0$ & $14.0 \pm 1.3$ & $13.7 \pm 1.3$ \\
6 & 1.30 & 1.36 & $18.8 \pm 1.2$ & $18.0 \pm 1.1$ & $18.6 \pm 1.4$ & $18.6 \pm 1.5$ \\
7 & 1.36 & 1.42 & $21.9 \pm 1.2$ & $23.0 \pm 1.3$ & $19.6 \pm 1.6$ & $19.7 \pm 1.6$ \\
8 & 1.42 & 1.48 & $25.5 \pm 1.4$ & $25.0 \pm 1.5$ & $25.5 \pm 1.7$ & $25.5 \pm 1.8$ \\
9 & 1.48 & 1.54 & $29.1 \pm 1.7$ & $27.3 \pm 1.9$ & $28.1 \pm 1.8$ & $26.0 \pm 1.9$ \\
10 & 1.54 & $w_{\mathrm{max}}$ & $32.2 \pm 2.5$ & $25.3 \pm 2.6$ & $30.5 \pm 2.3$ & $27.2 \pm 2.5$ \\
\bottomrule \bottomrule

\end{tabular}
\end{table*}

The fitted distributions in $(w, \cos\theta_{BY})$ bins are shown in Fig.~\ref{fig:postfit} for each of the four samples. The lower panel indicates the compatibility of the data with the post-fit model normalized by the statistical uncertainty. The goodness of fit, measured using $\chi^2$ per degree of freedom, is $335.4 / 318 = 1.05$, corresponding to a $p$-value of $24\%$.

The resulting central values of $\Delta\Gamma/\Delta w$, their uncertainties and their correlation coefficients are shown in Table~\ref{tab:dGdw}. The total fitted signal yields are listed in Table~\ref{tab:BR}. The individual differential decay rates in each subsample are shown in Table~\ref{tab:dGdwsubsamples}. The numerical values with full precision and the full covariance matrices are available on HEPData~\cite{hepdata.153613}.

The signal branching fractions are obtained in each sub-sample by summing the differential decay widths over \( w \), multiplied by the bin widths and the corresponding \( B \) meson lifetime. Uncertainties and correlations among the yields are propagated using multivariate Gaussian resampling based on the full covariance matrix. After averaging over lepton flavors, we find
\begin{equation}
  \mathcal{B}(B^+\to\bar D^0\ell^+\nu_\ell)=\resBrB,
\end{equation}
and
\begin{equation}
  \mathcal{B}(B^0\to D^-\ell^+\nu_\ell)=\resBrBz,
\end{equation}
which are consistent with the current world averages~\cite{HeavyFlavorAveragingGroupHFLAV:2024ctg}. The branching fractions in the subsamples are listed in Table~\ref{tab:BR}.

To test lepton flavor universality, we compute
\begin{equation}
  R^{e/\mu} = \mathcal{B}(B \to D e \nu_\ell)/\mathcal{B}(B \to D \mu \nu_\ell),
\end{equation}
where the branching fractions from the two \( B \) meson species are combined using their respective lifetimes. We find
\begin{equation}
  R^{e/\mu} = 1.020 \pm 0.020\,(\mathrm{stat.}) \pm 0.022\,(\mathrm{sys.}),
\end{equation}
consistent with the SM expectation.

\section{\texorpdfstring{Determination of $|V_{cb}|$}{Determination of Vcb}}
 \label{sec:vcb}

\subsection{BCL form factor fit} \label{sec:bcl}

\begin{table}
\caption{Parameters of the $N=3$ BCL expansion of the $f_+(q^2)$ and $f_0(q^2)$ form factor functions obtained from the fit described in Sec.~\ref{sec:bcl}.}
\begin{tabular}{p{1cm}p{1.6cm}*{6}{>{\centering\arraybackslash}p{0.8cm}}}
\toprule \toprule
& Values & \multicolumn{6}{c}{Correlation coefficients} \\
\midrule
$a_0$ & $\phantom{-}0.8959(92)$   &         & $1$                 & $\phantom{-}0.26$ & $-0.38$             & $\phantom{-}0.95$ & $\phantom{-}0.51$ \\
$a_1$ & $-8.03(15)$     &         &                     & $1$               & $\phantom{-}0.17$ & $\phantom{-}0.33$ & $\phantom{-}0.86$ \\
$a_2$ & $\phantom{-}49.3(3.1)$     &         &                     &                   & $1$               & $-0.31$             & $\phantom{-}0.16$ \\
$b_0$ & $\phantom{-}0.7813(73)$    &         &                     &                   &                   & $1$               & $\phantom{-}0.47$ \\
$b_1$ & $-3.38(15)$     &         &                     &                   &                   &                   & $1$ \\
\bottomrule \bottomrule
\end{tabular}
\label{tab:bcl}
\end{table}

The BCL expansion of the vector and scalar form factors, $f_+(q^2)$ and $f_0(q^2)$, truncated at $N=N^+=N^0=3$ has five free parameters (Eqs.~\eqref{eq:BCL1}, \eqref{eq:BCL2}),
\begin{equation}
    (c_i)=(a_0, a_1, a_2, b_0, b_1)~,
\end{equation}
as the parameter $b_2$ can be obtained from the kinematic constraint at $w_\mathrm{max}$ (Eq.~\eqref{eq:kinematic}). FLAG~\cite{FlavourLatticeAveragingGroupFLAG:2024oxs} has obtained the $B\to D$ form factor functions $f_+(q^2)$ and $f_0(q^2)$ precisely in this parametrization by averaging the calculations from FNAL/MILC~\cite{Lattice:2015rga} and HPQCD~\cite{Na:2015kha}. In this expansion, $t_+$ is set to $(m_{B^0}+m_{D^+})^2= 51.07$~GeV$^2$ and $t_0=(m_{B^0}+m_{D^+})(\sqrt{m_{B^0}}-\sqrt{m_{D^+}})^2 =  6.20$~GeV$^2$. We fit our result for $\Delta\Gamma_i/\Delta w$ (Table~\ref{tab:dGdw}) to this form factor calculation by minimizing
\begin{equation}
    \begin{split}
        \chi^2 = ~& \sum_{i,j}^{10}\left(\frac{\Delta\Gamma_i}{\Delta w}-\frac{\Delta\Gamma_{i,\mathrm{BCL}}}{\Delta w}\right)\mathbf{C}^{-1}_{ij}\left(\frac{\Delta\Gamma_j}{\Delta w}-\frac{\Delta\Gamma_{j,\mathrm{BCL}}}{\Delta w}\right) \\
        & + \sum_{k,l}^5\left(c_k-c_{k,\mathrm{FLAG}}\right)\mathbf{D}^{-1}_{kl}
        \left(c_l-c_{l,\mathrm {FLAG}}\right)~.
    \end{split}
\end{equation}
Here, $\Delta\Gamma_i/\Delta w$ are the measured values from Table~\ref{tab:dGdw} and $\Delta\Gamma_{i,\mathrm{BCL}}/\Delta w$ are the partial widths calculated using Eqs.~\eqref{eq:rate}, \eqref{eq:ff} and \ref{eq:BCL1}. The covariance matrix $\mathbf{C}_{ij}$ contains only experimental uncertainties. The FLAG calculation is contained in the parameter values $c_{k,\mathrm{FLAG}}$ and the theoretical covariance matrix $\mathbf{D}_{kl}$. The free parameters of the fit are the five BCL parameters $c_k$ and $\eta_\mathrm{EW}|V_{cb}|$ from Eq.~\eqref{eq:rate}.

\begin{figure}
    \centering
    \includegraphics[width=\columnwidth]{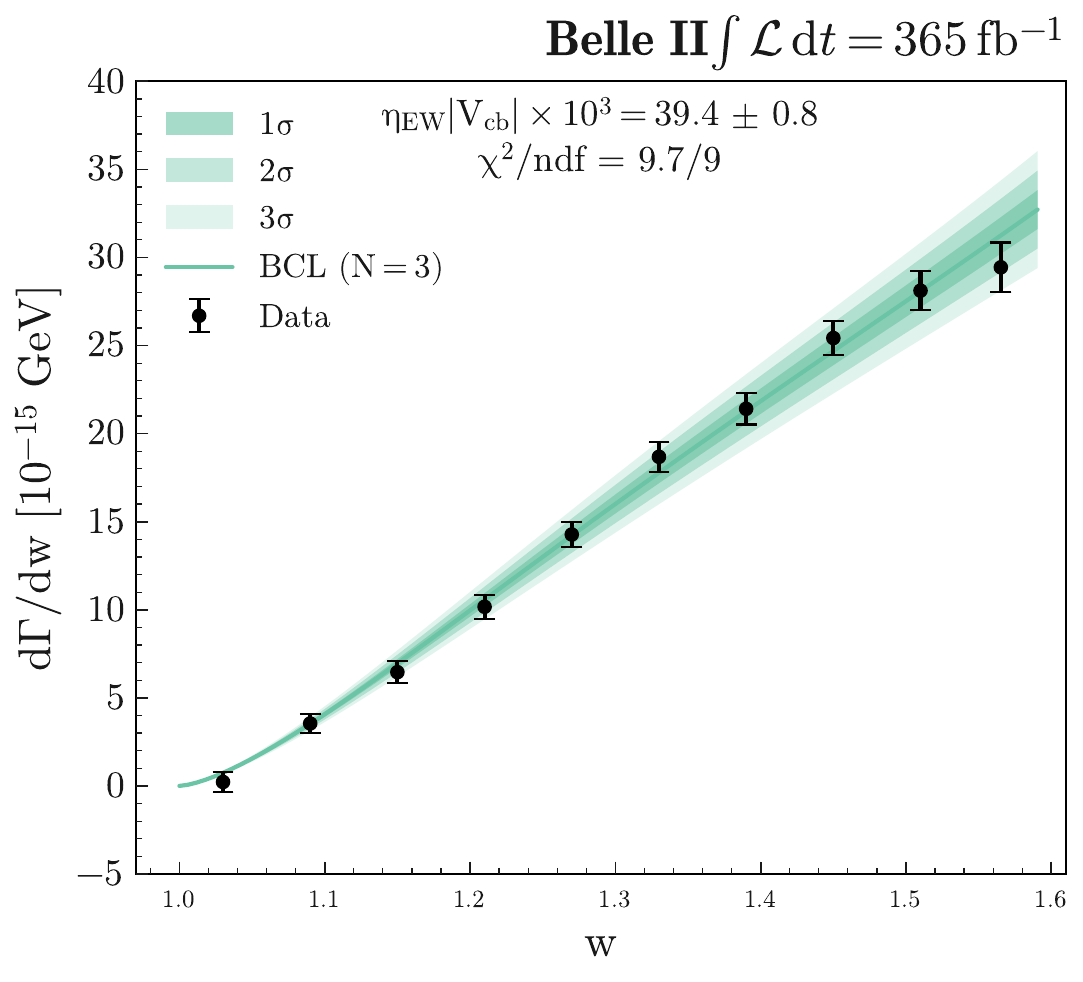}
    \caption{Form factor resulting from the fit described in Sec.~\ref{sec:bcl} compared to the measurement of $\Delta\Gamma_i/\Delta w$ (Table~\ref{tab:dGdw}). The uncertainty on $\eta_{\rm EW}|V_{cb}|$ shown represents the total uncertainty from statistical, systematic, and theoretical sources, as directly obtained from the fit, prior to decomposition into individual components.} \label{fig:bcl}
\end{figure}

Figure~\ref{fig:bcl} shows the result of our fit to the BCL expansion. The form factor parameters are given in Table~\ref{tab:bcl} together with their correlation coefficients. From the fit we obtain
\begin{equation}
\eta_{\rm EW} |V_{cb}|_\mathrm{BCL} = (39.4 \pm 0.8) \times 10^{-3}~,
\end{equation}
where the uncertainty includes all statistical, systematic and theoretical contributions. We split the uncertainty up into individual sources by resampling related nuisance parameters in the fit and varying the FLAG inputs within their uncertainty. The decomposition yields
\begin{equation}
\label{eq:vcbbclresult}
|V_{cb}|_\mathrm{BCL}=\resVcbBCL~.
\end{equation}
The value of the $\chi^2$~function at minimum is 9.7 for 9 degrees of freedom. The individual contributions to the uncertainty in $|V_{cb}|$ are listed in Table~\ref{tab:Vcb_syst}. A bin-by-bin extraction of
$|V_{cb}|$ and a comparison of the fitted form factors $f_+(w)$ and $f_0(w)$ to the lattice data are shown in the Supplemental Material~\cite{supplemental}.

\begin{table}
\caption{Fractional contributions to the total uncertainty on the extracted value of $|V_{cb}|$. The sizes of the contributions are given relative to the central value.} \label{tab:Vcb_syst}
\begin{tabular}{llr}
\toprule\toprule
\multicolumn{2}{c}{Source} & Uncertainty [\%] \\
\midrule
Statistical &                                           &   0.9 \\[1ex]
 Systematic &                                           &   1.5 \\[1ex]
            &                      $B^{0/+}$ lifetime   &   0.1 \\
            &                        Signal form factor &   0.1 \\
            &  $B \rightarrow D^* \ell \nu$ form factor &   0.1 \\
            & $\mathcal{B}(B \rightarrow X_c \ell \nu)$ &   0.3 \\
            &  $\mathcal{B}(D \rightarrow K \pi (\pi))$ &   0.5 \\
            &                       Tracking efficiency &   0.5 \\
            &                        $N_{\Upsilon(4S)}$ &   0.7 \\
            &                           $f_{00}/f_{+-}$ &   0.1 \\
            &                        $f_{\slashed{B}}$  &   0.4 \\
            &                 Background $w$ modelling  &   0.3 \\
            &                  $(E^*_Y,m_Y)$ reweighting &   0.3 \\
            &                     Lepton identification &   0.3 \\
            &                       Kaon identification &   0.6 \\
            &          Vertex fit $\chi^2$ correction   &   0.3 \\
            &                    Simulation sample size &   0.5 \\[1ex]
Theoretical &                                           &   1.3 \\[1ex]
            &                        Lattice QCD inputs &   1.2 \\
            &                         Long-distance QED &   0.5 \\[1ex]
      Total &                                           &   2.1 \\[1ex]
\bottomrule\bottomrule
\end{tabular}

\end{table}

\subsection{CLN form factor fit} \label{subsec:CLN}

We also use the measured $\Delta\Gamma_i/\Delta w$ spectrum averaged over the four modes to perform a fit to the differential decay rate Eq.~\eqref{eq:rate} assuming the CLN~form factor Eq.~\eqref{eq:CLN} and determine the form factor at zero recoil times the CKM matrix element magnitude $\eta_\mathrm{EW}\mathcal{G}(1)|V_{cb}|$ and the form factor slope parameter $\rho^2$. This is a least-squares fit with the $\chi^2$~function
\begin{equation}
  \chi^2 = \sum\limits_{i,j}\left(\frac{\Delta\Gamma_i}{\Delta w}-\frac{\Delta\Gamma_{i,\mathrm{CLN}}}{\Delta w}\right)\mathbf{C}^{-1}_{ij}\left(\frac{\Delta\Gamma_j}{\Delta w}-\frac{\Delta\Gamma_{j,\mathrm{CLN}}}{\Delta w}\right)~,
\end{equation}
where $\Delta\Gamma_i/\Delta w$ are the measured values from Table~\ref{tab:dGdw} and $\Delta\Gamma_{i,\mathrm{CLN}}/\Delta w$ are the partial widths calculated using Eqs.~\eqref{eq:rate} and \eqref{eq:CLN}. We use the averaged masses of charged and neutral mesons. 

The result of the fit is
\begin{eqnarray}
 \eta_\mathrm{EW}\mathcal{G}(1)|V_{cb}| & = & (40.9  \pm  1.4)\times 10^{-3}~, \\
 \rho^2 & = & 1.09 \pm 0.06~,
\end{eqnarray}
with a correlation $\rho_{\eta_\mathrm{EW}\mathcal{G}(1)|V_{cb}|,\rho^2}=0.89$. The uncertainties and the correlation coefficient include only experimental contributions. The $\chi^2$ of the fit is 5.9 for 8 degrees of freedom. Using $\mathcal{G}(1)=1.0541\pm 0.0083$~\cite{Lattice:2015rga}, we find
\begin{equation}
|V_{cb}|_\mathrm{CLN}=\resVcbCLN~.
\end{equation}
The results for $\eta_\mathrm{EW}\mathcal{G}(1)|V_{cb}|$ and $\rho^2$ are consistent with previous measurements by BaBar and Belle~\cite{BaBar:2009zxk,Belle:2015pkj}. Due to the limited precision of the CLN form factor (Sec.~\ref{sec:theo}) we select $|V_{cb}|$ obtained with the BCL form factor (Sec.~\ref{sec:bcl} as our central value.

\section{Summary} \label{sec:Summary}

We reconstruct about 87\,000 $B^0\to D^-\ell^+\nu_\ell$ decays and about 136\,000 $B^+\to\bar D^0\ell^+\nu_\ell$ decays in \lumi of $e^+e^-\to\Upsilon(4S)\to B\bar B$ data recorded with the Belle~II experiment at the SuperKEKB collider. The branching fractions of these decays are found to be
\begin{eqnarray}
  \mathcal{B}(B^0\to D^-\ell^+\nu_\ell) & = & \resBrBz~, \nonumber \\
  \mathcal{B}(B^+\to\bar D^0\ell^+\nu_\ell) & = & \resBrB~. \nonumber
\end{eqnarray}
We also probe lepton flavor universality in the $b\to c$~weak transition and determine

\begin{equation}
\begin{split}
\mathcal{B}(B\to D e\nu_e)/\mathcal{B}(B\to D\mu\nu_\mu) = {} & 1.020 \pm 0.020\,(\mathrm{stat.}) \\
                   & \pm 0.022\,(\mathrm{sys.})  \nonumber
\end{split}
\end{equation}

By using inclusive reconstruction of the unobserved neutrino momentum, we measure the recoil variable $w=v_B\cdot v_D$ in every $B\to D\ell\nu_\ell$ signal event, where $v_B$ and $v_D$ are the four-velocities of the $B$ and $D$~mesons, respectively. This allows the determination of the differential $B\to D\ell\nu_\ell$ width, $\Delta\Gamma_i/\Delta w$, in ten bins of $w$, given in Table~\ref{tab:dGdw} and on HEPData~\cite{hepdata.153613}. We determine the magnitude of the CKM matrix element $V_{cb}$ by fitting our measurement of $\Delta\Gamma_i/\Delta w$ to the theoretical expression of the rate assuming the BCL expansion of the form factor~\cite{Bourrely:2008za}. In this fit, the coefficients of the BCL expansion are constrained to the FLAG average~\cite{FlavourLatticeAveragingGroupFLAG:2024oxs} of the FNAL/MILC~\cite{Lattice:2015rga} and HPQCD~\cite{Na:2015kha} calculations of the $B\to D\ell\nu_\ell$ form factor. For $|V_{cb}|$, we obtain
\begin{equation}
|V_{cb}|=\resVcbBCL~, \nonumber
\end{equation}
where the uncertainty is separated into statistical, systematic and theoretical contributions.

This determination of the CKM matrix element $|V_{cb}|$ is consistent with previous results from BaBar and Belle~\cite{BaBar:2009zxk,babarcollaboration2023modelindependentextractionformfactors,Belle:2015pkj} but with 2.1\% total uncertainty significantly improves the precision of $|V_{cb}|$ from the decay $B\to D\ell\nu_\ell$. The resulting value of $|V_{cb}|$ is also in agreement with the value obtained by HFLAV from a global fit to exclusive measurements $|V_{cb}| = (39.62 \pm 0.47) \times 10^{-3}$~\cite{HeavyFlavorAveragingGroupHFLAV:2024ctg}.

\section*{Acknowledgments}
This work, based on data collected using the Belle II detector, which was built and commissioned prior to March 2019,
was supported by
Higher Education and Science Committee of the Republic of Armenia Grant No.~23LCG-1C011;
Australian Research Council and Research Grants
No.~DP200101792, 
No.~DP210101900, 
No.~DP210102831, 
No.~DE220100462, 
No.~LE210100098, 
and
No.~LE230100085; 
Austrian Federal Ministry of Education, Science and Research,
Austrian Science Fund (FWF) Grants
DOI:~10.55776/P34529,
DOI:~10.55776/J4731,
DOI:~10.55776/J4625,
DOI:~10.55776/M3153,
and
DOI:~10.55776/PAT1836324,
and
Horizon 2020 ERC Starting Grant No.~947006 ``InterLeptons'';
Natural Sciences and Engineering Research Council of Canada, Compute Canada and CANARIE;
National Key R\&D Program of China under Contract No.~2024YFA1610503,
and
No.~2024YFA1610504
National Natural Science Foundation of China and Research Grants
No.~11575017,
No.~11761141009,
No.~11705209,
No.~11975076,
No.~12135005,
No.~12150004,
No.~12161141008,
No.~12475093,
and
No.~12175041,
and Shandong Provincial Natural Science Foundation Project~ZR2022JQ02;
the Czech Science Foundation Grant No. 22-18469S,  Regional funds of EU/MEYS: OPJAK
FORTE CZ.02.01.01/00/22\_008/0004632
and
Charles University Grant Agency project No. 246122;
European Research Council, Seventh Framework PIEF-GA-2013-622527,
Horizon 2020 ERC-Advanced Grants No.~267104 and No.~884719,
Horizon 2020 ERC-Consolidator Grant No.~819127,
Horizon 2020 Marie Sklodowska-Curie Grant Agreement No.~700525 ``NIOBE''
and
No.~101026516,
and
Horizon 2020 Marie Sklodowska-Curie RISE project JENNIFER2 Grant Agreement No.~822070 (European grants);
L'Institut National de Physique Nucl\'{e}aire et de Physique des Particules (IN2P3) du CNRS
and
L'Agence Nationale de la Recherche (ANR) under Grant No.~ANR-21-CE31-0009 (France);
BMFTR, DFG, HGF, MPG, and AvH Foundation (Germany);
Department of Atomic Energy under Project Identification No.~RTI 4002,
Department of Science and Technology,
and
UPES SEED funding programs
No.~UPES/R\&D-SEED-INFRA/17052023/01 and
No.~UPES/R\&D-SOE/20062022/06 (India);
Israel Science Foundation Grant No.~2476/17,
U.S.-Israel Binational Science Foundation Grant No.~2016113, and
Israel Ministry of Science Grant No.~3-16543;
Istituto Nazionale di Fisica Nucleare and the Research Grants BELLE2,
and
the ICSC – Centro Nazionale di Ricerca in High Performance Computing, Big Data and Quantum Computing, funded by European Union – NextGenerationEU;
Japan Society for the Promotion of Science, Grant-in-Aid for Scientific Research Grants
No.~16H03968,
No.~16H03993,
No.~16H06492,
No.~16K05323,
No.~17H01133,
No.~17H05405,
No.~18K03621,
No.~18H03710,
No.~18H05226,
No.~19H00682, 
No.~20H05850,
No.~20H05858,
No.~22H00144,
No.~22K14056,
No.~22K21347,
No.~23H05433,
No.~26220706,
and
No.~26400255,
and
the Ministry of Education, Culture, Sports, Science, and Technology (MEXT) of Japan;
National Research Foundation (NRF) of Korea Grants
No.~2021R1-F1A-1064008,
No.~2022R1-A2C-1003993,
No.~2022R1-A2C-1092335,
No.~RS-2016-NR017151,
No.~RS-2018-NR031074,
No.~RS-2021-NR060129,
No.~RS-2023-00208693,
No.~RS-2024-00354342
and
No.~RS-2025-02219521,
Radiation Science Research Institute,
Foreign Large-Size Research Facility Application Supporting project,
the Global Science Experimental Data Hub Center, the Korea Institute of Science and
Technology Information (K25L2M2C3 )
and
KREONET/GLORIAD;
Universiti Malaya RU grant, Akademi Sains Malaysia, and Ministry of Education Malaysia;
Frontiers of Science Program Contracts
No.~FOINS-296,
No.~CB-221329,
No.~CB-236394,
No.~CB-254409,
and
No.~CB-180023, and SEP-CINVESTAV Research Grant No.~237 (Mexico);
the Polish Ministry of Science and Higher Education and the National Science Center;
the Ministry of Science and Higher Education of the Russian Federation
and
the HSE University Basic Research Program, Moscow;
University of Tabuk Research Grants
No.~S-0256-1438 and No.~S-0280-1439 (Saudi Arabia), and
Researchers Supporting Project number (RSPD2025R873), King Saud University, Riyadh,
Saudi Arabia;
Slovenian Research Agency and Research Grants
No.~J1-50010
and
No.~P1-0135;
Ikerbasque, Basque Foundation for Science,
State Agency for Research of the Spanish Ministry of Science and Innovation through Grant No. PID2022-136510NB-C33, Spain,
Agencia Estatal de Investigacion, Spain
Grant No.~RYC2020-029875-I
and
Generalitat Valenciana, Spain
Grant No.~CIDEGENT/2018/020;
The Knut and Alice Wallenberg Foundation (Sweden), Contracts No.~2021.0174 and No.~2021.0299;
National Science and Technology Council,
and
Ministry of Education (Taiwan);
Thailand Center of Excellence in Physics;
TUBITAK ULAKBIM (Turkey);
National Research Foundation of Ukraine, Project No.~2020.02/0257,
and
Ministry of Education and Science of Ukraine;
the U.S. National Science Foundation and Research Grants
No.~PHY-1913789 
and
No.~PHY-2111604, 
and the U.S. Department of Energy and Research Awards
No.~DE-AC06-76RLO1830, 
No.~DE-SC0007983, 
No.~DE-SC0009824, 
No.~DE-SC0009973, 
No.~DE-SC0010007, 
No.~DE-SC0010073, 
No.~DE-SC0010118, 
No.~DE-SC0010504, 
No.~DE-SC0011784, 
No.~DE-SC0012704, 
No.~DE-SC0019230, 
No.~DE-SC0021274, 
No.~DE-SC0021616, 
No.~DE-SC0022350, 
No.~DE-SC0023470; 
and
the Vietnam Academy of Science and Technology (VAST) under Grants
No.~NVCC.05.12/22-23
and
No.~DL0000.02/24-25.

These acknowledgements are not to be interpreted as an endorsement of any statement made
by any of our institutes, funding agencies, governments, or their representatives.

We thank the SuperKEKB team for delivering high-luminosity collisions;
the KEK cryogenics group for the efficient operation of the detector solenoid magnet and IBBelle on site;
the KEK Computer Research Center for on-site computing support; the NII for SINET6 network support;
and the raw-data centers hosted by BNL, DESY, GridKa, IN2P3, INFN,
and the University of Victoria.

\bibliography{references}

\end{document}